\documentclass[letterpaper,10pt]{article} 
\usepackage{osameet3} 
\usepackage{wrapfig}
\usepackage{amsmath,amssymb}
\usepackage[colorlinks=true,bookmarks=false,citecolor=blue,urlcolor=blue]{hyperref}

\usepackage{tikz}
\usepackage{pgfplots}
    \pgfplotsset{compat=1.15}
    \usepgfplotslibrary{fillbetween}
    \usepackage{pgfplotstable}
    \usepackage{xstring}
\usepackage{subcaption}
\usepackage{graphicx}
\usepackage{amsmath}
\newcommand*\rel@kern[1]{\kern#1\dimexpr\macc@kerna}
\newcommand*\widebar[1]{%
  \begingroup
  \def\mathaccent##1##2{%
    \rel@kern{0.8}%
    \overline{\rel@kern{-0.8}\macc@nucleus\rel@kern{0.2}}%
    \rel@kern{-0.2}%
  }
  \macc@depth\@ne
  \let\math@bgroup\@empty \let\math@egroup\macc@set@skewchar
  \mathsurround\z@ \frozen@everymath{\mathgroup\macc@group\relax}%
  \macc@set@skewchar\relax
  \let\mathaccentV\macc@nested@a
  \macc@nested@a\relax111{#1}%
  \endgroup
}
\DeclareMathAlphabet{\pazocal}{OMS}{zplm}{m}{n}
\SetMathAlphabet\pazocal{bold}{OMS}{zplm}{bx}{n}

\begin{document}

\title{Temporal Properties of Enumerative Shaping: Autocorrelation and Energy Dispersion Index}

\author{
Yunus Can Gültekin, Kaiquan Wu, and Alex Alvarado
}
\address{{Department of Electrical Engineering, Eindhoven University of Technology, The Netherlands.} \\
y.c.g.gultekin@tue.nl}

\vspace{-0.4cm}

\copyrightyear{2022}

\begin{abstract}
We study the effective SNR behavior of various enumerative amplitude shaping algorithms.
We show that their relative behavior can be explained via the temporal autocorrelation function or via the energy dispersion index.
\end{abstract}

\vspace{-0.1cm}

\section{Introduction}

Probabilistic amplitude shaping (PAS) has become an indispensable technique for optical communication systems in recent years~\cite{Bocherer2015,Fehenberger2016}.
PAS provides signal-to-noise ratio (SNR) gains more than 1 dB over uniform quadrature amplitude modulation (QAM) constellations and enables rate adaptivity.
Thanks to these properties, PAS has been numerically and experimentally demonstrated to yield more than 30\% reach increase for a wide range of transmission rates~\cite{Buchali2016,Goossens2019,Amari2019}.
To secure these gains with low complexity, recently-developed enumerative coding algorithms such as enumerative sphere shaping (ESS), kurtosis-limited ESS (K-ESS), and band-trellis ESS (B-ESS) have been widely investigated as efficient amplitude shaping techniques~\cite{Gultekin2020,Amari20192,Gultekin2021,Gultekin2022,Cho2022}.
It has been observed that they offer different effective SNRs.
Moreover, ESS and K-ESS result in SNR penalties against uniform signaling, while B-ESS provides SNR gains.
In this work, we seek to understand why this is the case.
We study enumerative shaping techniques via two metrics that characterize temporal properties of channel inputs: auto-correlation function (ACF) of~\cite[Ch. 6]{Fehenberger2017} and recently-introduced energy dispersion index (EDI) of~\cite[Defn. 5]{Wu2021}.
We show that B-ESS is the most efficient in both ACF- and EDI-sense, and this is the reason why it provides the largest SNR.

\vspace{-0.2cm}

\section{System Model: Probabilistic Amplitude Shaping}

PAS combines an amplitude shaper with a channel code.
Here, we focus on the shaping blocks implemented via enumerative coding techniques: ESS~\cite{Amari2019}, K-ESS~\cite{Gultekin2021}, and B-ESS~\cite{Gultekin2022}.
With ESS, amplitude $N$-sequences $\underline{a} = (a_1, a_2,\dotsc, a_N)$ that satisfy a maximum-energy constraint $\sum_{k=1}^N |a_k|^2 < E_{\max}$ are generated. 
With K-ESS, amplitude sequences {\it also} satisfy a maximum-kurtosis constraint $\sum_{k=1}^N |a_k|^4 < K_{\max}$.
Finally, with B-ESS, sequences have limited windowed energy variations.
Then the amplitude sequences are converted into a channel input sequence $\underline{x}$ of QAM symbols.
At the receiver, the channel output $\underline{y}$ is observed.
We want to find insights into the memory-related effects of the channel on $\underline{x}$.
In what follows, we discuss two metrics to study these effects.

In~\cite{Fehenberger2017}, the temporal correlations are described via the ACF $R_{\theta}[\tau]$ of the moving-window average angle $\theta_k$ of past $w$ symbols, see (1) and (2), resp.
Here, $w$ is the moving average window length, $\angle z$ is the phase of $z$, and $R_{\theta}[\tau]$ is normalized to have $R_{\theta}[0] = 1$.
The correlations $R_{\theta}[\tau]$ for $\tau\leq w$ are introduced by the moving average in (2).
After $w$ symbols, the correlations $R_{\theta}[\tau]$ give insight about the effective memory observed in the channel.

In~\cite{Wu2021}, the time-varying characteristics of the energy in the transmitted signal are quantified via the EDI $\Psi$ in (3).
EDI is the normalized average of the variations in windowed energy $G^W$ in (4).
Here, $W$ is the window length.
It has been shown in the literature that there are strong correlations between EDI and effective SNR~\cite{Wu2021,Wu2021_2}.
The optimum window length $W$---in the sense that the absolute value of the correlation between EDI and SNR is maximized---depends on the length of the link and gives an insight into the memory of the channel~\cite[Fig. 9]{Wu2021}.

\begin{equation}
    R_{\theta}[\tau] = \frac{E[\theta_{k}\theta_{k-\tau}]}{n-\tau}. \hspace{0.15cm} (1) \hspace{0.4cm} \theta_k = \angle \sum_{m=-(w-1)/2}^{(w-1)/2} \boldsymbol{x}^{*}_{k-m} \boldsymbol{y}_{k-m}. \hspace{0.15cm} (2) \hspace{0.4cm} \Psi = \frac{\widebar{\text{Var}}(G^W)}{\widebar{E}(G^W)}. \hspace{0.15cm} (3) \hspace{0.4cm} G^W_k = \sum_{i=k-W/2}^{k+W/2} |x_i|^2. \hspace{0.2cm} (4) \nonumber
\end{equation}

In the next section, we will provide numerical results showing that these two metrics, ACF in (1) and EDI in (3), can be used to predict the effective SNRs of shaped signaling schemes.

\vspace{-0.2cm}

\section{Results}

We simulate the transmission of a single-polarized single-channel waveform over 1 and 4 spans of 80 km fiber with an attenuation of 0.19 dB/km, a dispersion of 17 ps/nm/km, and a nonlinear (NL) parameter of 1.3 1/W/km using the split-step Fourier method (SSFM). 
A root-raised-cosine pulse with 10\% roll-off is used.
The symbol rate is 56 GBd.
The fiber is followed by an erbium-doped fiber amplifier with a noise figure of 5.5 dB. 
These parameters are compatible with the ones used in~\cite{Wu2021}.
At the receiver, first, the chromatic dispersion, then the phase rotation induced by the NL are ideally compensated.
PAS is realized based on ESS, K-ESS, and B-ESS with the parameters used to generate~\cite[Fig. 3]{Gultekin2022} with 64-QAM.
Shaping blocklength is $N=108$~amplitudes.
The shaping rate is 1.5 bit/1D, which leads to a net information rate of 8 bit/4D when combined with a rate-5/6 channel code.
However here, we do not realize channel coding and select the signs for the amplitudes uniformly at random.

\begin{figure}[ht]
    \makebox[\textwidth]{\makebox[1\textwidth]{
    \centering
    \begin{subfigure}[t]{.329\linewidth}%
        \centering 
	    \resizebox{\columnwidth}{!}{\includegraphics{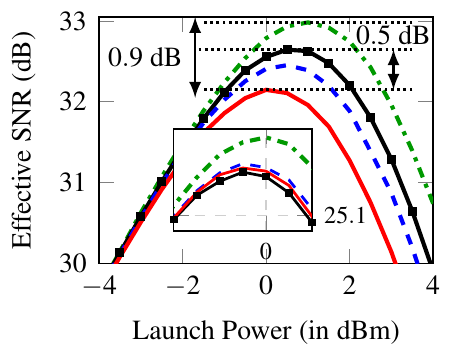}}	    
        \label{fig:SNR}%
    \end{subfigure}
    \begin{subfigure}[t]{.329\linewidth}
        \centering 
	    \resizebox{\columnwidth}{!}{\includegraphics{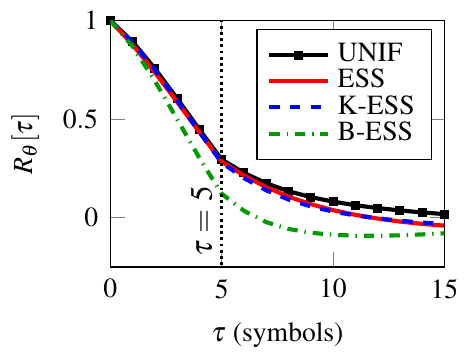}}	
        \label{fig:Autocorr}%
    \end{subfigure}
    \begin{subfigure}[t]{.329\linewidth}%
        \centering 
	    \resizebox{\columnwidth}{!}{\includegraphics{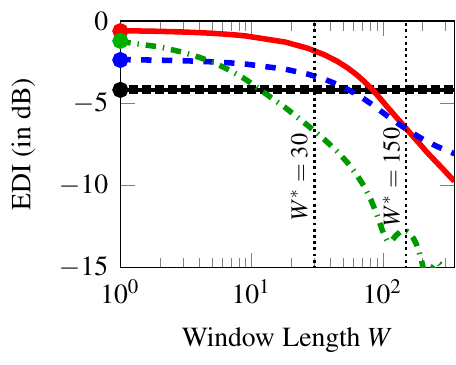}}	
        \label{fig:EDI}%
    \end{subfigure}
    }}
    \vspace{-7mm}
    \caption{{\bf Left:} Effective SNR after 1 span, {\bf inset:} after 4 spans. {\bf Middle:} ACF with $w=5$. {\bf Right:} EDI.\label{fig:results}\vspace{-5mm}}
\end{figure}

Fig.~\ref{fig:results} (left) shows the effective SNR vs. the launch power $P$ after 1 and 4 spans of transmission. 
In Fig.~\ref{fig:results} (middle), we show the ACF $R_\theta[\tau]$ as a function of $\tau$ for $w=5$.
This ACF is computed for input-output pairs belonging to the 1-span scenario at the corresponding optimum launch powers. 
We see that ESS and K-ESS result in ACFs very similar to uniform signaling (except at the tail ends).
On the other hand, B-ESS has an ACF function which vanishes more quickly.
This implies that when input sequences generated via B-ESS are transmitted, the observed effective memory is shorter than it is for ESS, K-ESS, and uniform signaling.
We believe this explains (at least partially) the significantly-improved end-to-end decoding performance provided by B-ESS~\cite[Figs. 3-4]{Gultekin2022} based on typical mismatched bit-metric decoding in which the memory in the channel is neglected.

In Fig.~\ref{fig:results} (right), we show the EDI of considered schemes as a function of the moving average length $W$.
We also indicate the optimum window lengths $W^{*}=30$ and $150$ for the considered setup after 1 and 4 spans of transmission~\cite[Figs. 9a-b]{Wu2021}, resp.
The relative behaviors of EDIs at $W=30$ and $150$ are the same as that of SNRs in Fig.~\ref{fig:results} (left): a smaller EDI results in a larger SNR.
Since B-ESS was designed specifically to limit the variations in channel input signal energy, it provides the smallest EDI for most of the $W$ region (especially around $W=N$).
This can also be used to explain the significant SNR improvement provided by B-ESS in Fig.~\ref{fig:results} (left).
Moreover, this correlation between EDI and SNR can be exploited to design B-ESS for different link setups without resorting to computationally-expensive SSFM-based simulations.

Finally, we have checked numerically and discovered that K-ESS has smaller average energy, smaller average energy variance, and smaller kurtosis than B-ESS, i.e., it is {\it better} in terms of average statistics which neglect temporal characteristics.
This implies that these statistics are insufficient to capture the properties of shaping schemes that influence the resulting NL interference and hence, the effective SNR.
EDI, on the other hand, focuses on temporal characteristics, evaluates windowed energy variations, and qualitatively predicts the SNR accurately.

\vspace{-0.2cm}

\section{Conclusions}
We have shown via fiber simulations that the relative effective SNR behaviors of uniform vs. enumeratively shaped signaling (via ESS, K-ESS, and B-ESS)  can be explained via arguments based on autocorrelation function (ACF) and energy dispersion index (EDI).
B-ESS---limiting windowed energy variations in input sequences---has the most quickly-vanishing ACF and the smallest EDI and hence, provides the largest SNR.

\vspace{-2mm}

\noindent\footnotesize\emph{\\
The work of Y.C. G\"{u}ltekin, K. Wu, and A. Alvarado has received funding from the NWO via the VIDI grant ICONIC (grant ID: 15685), and from the ERC under the EU's Horizon 2020 research and innovation programme via the Starting grant FUN-NOTCH (grant ID: 757791) and via the Proof of Concept grant SHY-FEC (grant ID: 963945).}

\end{document}